\documentclass[prd,twocolumn,showpacs,preprintnumbers,amsmath,amssymb,superscriptaddress,floatfix,nofootinbib]{revtex4}

\usepackage{graphicx}
\usepackage{epstopdf}
\usepackage{bm}
\usepackage{amsmath}
\usepackage{amsfonts}
\usepackage{amssymb}
\usepackage{color}
\usepackage{multirow}
\usepackage{footnote}

\usepackage[colorlinks, citecolor=blue,anchorcolor=red,menucolor=red,linkcolor=red,filecolor=red,runcolor=red,urlcolor=blue,frenchlinks=red]{hyperref}
\begin{document}

\title{The assignments of the $B_s$ mesons within the screened potential model and $^3P_0$ model}

\author{Wei Hao}\email{haowei2020@itp.ac.cn}
\affiliation{School of Physics and Microelectronics, Zhengzhou University, Zhengzhou, Henan 450001, China}
\affiliation{CAS Key Laboratory of Theoretical Physics, Institute of Theoretical Physics, Chinese Academy of Sciences, Beijing 100190,China}
\affiliation{University of Chinese Academy of Sciences (UCAS), Beijing 100049, China}

\author{Yu Lu}
\email{ylu@ucas.ac.cn}
\affiliation{School of Physical Sciences, University of Chinese Academy of Sciences (UCAS), Beijing 100049, China}

\author{En Wang}\email{wangen@zzu.edu.cn}
\affiliation{School of Physics and Microelectronics, Zhengzhou University, Zhengzhou, Henan 450001, China}

\begin{abstract}
We investigate the mass spectrum and the decay properties of the $B_s$ mesons within the screened nonrelativistic quark model and the $^3P_0$ model. Our results suggest that the $B_{sJ}(6064)$ and $B_{sJ}(6114)$ states, as the first solution of the recently LHCb measurements, could be explained as the  $B_s(1^3D_3)$ and $B_s(1^3D_1)$, respectively. In addition, the $B_{sJ}(6109)$ and $B_{sJ}(6158)$ states, as the second solution of the LHCb measurements, could be explained as the $B'_{s2}(1D)$ 
and $B_{s1}(2P)$, respectively. Meanwhile,  the $B_{s1}(5830)$ could be interpreted as the candidate of the $B_{s1}(1P)$.  
We also calculated the decay properties of the other excited $B_s$ mesons with the predicted masses, which should be helpful for the experimental searching  in  future.
\end{abstract}

\maketitle

\section{Introduction}

Recently, the LHCb Collaboration observed an excess structure 300~MeV above the $B^\pm K^\mp$ threshold in the $B^\pm K^\mp$ mass spectrum in the proton-proton collisions, which could be described by a two-peak
hypothesis~\cite{LHCb:2020pet}. 
Assuming they decay directly to the $B^\pm K^\mp$ final states, the two peaks could be associated to the resonances $B_{sJ}(6064)$ and $B_{sJ}(6114)$, with the masses and widths as follows,
\begin{eqnarray}
    \label{exp1}
    B_{sJ}(6064): M &=& 6063.5\pm1.2\pm0.8~{\rm MeV},\\
             \Gamma &=& 26\pm4\pm4~{\rm MeV};\\
    B_{sJ}(6114): M &=& 6114\pm3\pm5~{\rm MeV},\\
             \Gamma &=& 66\pm18\pm21~{\rm MeV}.
 \end{eqnarray}
However, if a decay through $B^{*\pm}K^\mp$ with a missing photon from the $B^{*\pm} \to B^\pm\gamma$ decay is assumed, the masses and widths will shift to be,
\begin{eqnarray}
    \label{exp2}
    B_{sJ}(6109): M &=& 6108.8\pm1.1\pm0.7~{\rm MeV},\\
             \Gamma &=& 22\pm5\pm4~{\rm MeV};\\
    B_{sJ}(6158): M &=& 6158\pm4\pm5~{\rm MeV},\\
             \Gamma &=& 72\pm18\pm25~{\rm MeV} .
 \end{eqnarray}
 These observations of the LHCb have enriched the bottom-strange spectrum. We have tabulated the experimental information of all the $B_s$ mesons in Table~\ref{experment}.

Although these two states were suggested to be the $D$-wave orbital excited bottom-strange mesons,  their masses are significantly lower than the
quark model predictions~\citep{Lu:2016bbk,Godfrey:2016nwn,Ebert:2009ua,Sun:2014wea}. There have been  some theoretical works to study these states~\citep{Chen:2022fye,li:2021hss,Patel:2022hhl,Gandhi:2022nnk} (A recent review about these two states can be found in Ref.~\cite{Chen:2022asf}). Based on the nonrelativistic linear potential model, Ref.~\citep{li:2021hss} explained the $B_{sJ}(6114)$ as the $B_s(2^3S_1)-B_s(1^3D_1)$ mixing state, and obtained the mass $M=6114$~MeV and width $\Gamma=95\pm15$~MeV with the mixing angle $\theta=-(45\pm16)^\circ$, which is  supported by the results based on the screened potential model~\citep{Patel:2022hhl}. In addition, Ref.~\citep{li:2021hss} also supports the interpretation of the $B_{sJ}(6114)$ as a pure $B_s(1^3D_1)$ state if there is small mixing between $B_s(2^3S_1)$ and $B_s(1^3D_1)$.

For the $B_{sJ}(6064)$, there are also two possible interpretations in Ref.~\citep{li:2021hss}. According to the first interpretation, the narrow structure around 6064~MeV is mainly caused by the $B_{sJ}(6109)$ resonance, which is regarded as the $B_s(1^1D_2)-B_s(1^3D_2)$ mixing state. For the second interpretation, the $B_{sJ}(6064)$ could be explained as a pure $B_s(1^3D_3)$ with the predicted mass $M=6067$~MeV and width  $\Gamma=13$~MeV.

In addition, authors of Ref. \citep{Gandhi:2022nnk} have studied the $B$ and $B_s$ mesons using the heavy quark effective theory, which suggests that the $B_{sJ}(6064)$ could be the candidate of the $B_s(2^3S_1)$ with the predicted  mass $6033.0\pm2.4$~MeV and width $170\pm1.5$~MeV, and the $B_{sJ}(6114)$ could be the candidate of the $B_s(1^3D_3)$ with predicted mass $6247.0\pm2.4$~MeV and width $82.0\pm1.0$~MeV. However, the predicted width for $B_{sJ}(6064)$ and mass for $B_{sJ}(6114)$ are larger than experimental values.

Besides the conventional $B_s$ mesons explanation, the two states are also regarded as $b\bar{s}q\bar{q}$ tetraquark states in Ref.~\citep{Chen:2021uou}, or as the $\bar{B}K^*$ molecular state with quantum numbers $(I)J^P=(0)1^+$ in Ref.~\citep{Kong:2021ohg}.
Thus, one can find that the natures of these two states are still in debate, and more efforts are needed to shed light on their internal structures.

As we known, although the quenched quark models have obtained lots of success in the last decades, the effects of the sea quarks and gluons interactions are not taken into account. The unquenched models, considering all kinds of additional effects, have been developed, and widely used to  describe the hadron spectra,  such as the coupled channel model~\citep{Ferretti:2013faa,Ferretti:2013vua,Ortega:2009hj,Ortega:2016mms,Hao:2022vwt,Yang:2023tvc,Xie:2021dwe} and the screened potential model \citep{Song:2015fha,Song:2015nia,Wang:2018rjg,Wang:2019mhs,Hao:2019fjg,Feng:2022esz}. In the quenched potential model, the  potentials mainly contain a coulomb term at short distances and the linear confining interaction at large distances. However this is not appreciate in the large mass range, since the linear potential, which is expected to be dominant in large mass region, will be
screened or softened by the vacuum polarization effects of dynamical fermions~\citep{Born:1989iv,Li:2009zu}, {\sl i.e.}, the unquenched effects reflecting the sea quarks or gluons contributions to some extent.
Clearly, the unquenched effects can lead to important influence for higher radial and orbital excited hadrons, which means that the predicted masses of the higher excited states  will be smaller than the ones of the general liner potential models. Comparing with the coupled channel model, the screened potential model is simpler, and have been successfully used to describe the spectra of the charmed-strange meson~\citep{Song:2015nia,Gao:2022bsb}, charmed meson~ \citep{Song:2015fha}, excited $\rho$ mesons~\citep{Feng:2022hwq,Li:2021qgz}, bottom mesons~\cite{Feng:2022esz}, charmonium~\citep{Wang:2019mhs,Li:2009zu,Hao:2019fjg}, and bottomonium~\citep{Wang:2018rjg,Li:2009nr}.


\begin{table}[!htpb]
\begin{center}
\caption{ \label{experment} Experimental information of the $B_s$ mesons~\citep{ParticleDataGroup:2022pth}.}
\footnotesize
\setlength{\tabcolsep}{1mm}{
\begin{tabular}{ccccc}
\hline\hline
  state                 &mass  (MeV)                          &width  (MeV)                & $I(J^{P})$          \\\hline
  $B_s$               &$5366.92\pm0.10$                     &$-$                   & $0(0^-)$  \\ 
  $B_s^*$               &$5415.4^{+1.8}_{-1.5}$               &$-$                   & $0(1^-)$      \\
  $B_{s1}(5830)$      &$5828.70\pm 0.20$                    &$0.5\pm0.3\pm 0.3$           & $0(1^+)$ \\
  $B_{s2}^*(5840)$    &$5839.86\pm0.12$                     &$1.49\pm0.27$         & $0(2^+)$    \\
  $B_{sJ}(6064)$      &$6063.5\pm1.2\pm0.8$                 &$26\pm4\pm4$          & $0(?^?)$\cite{LHCb:2020pet}     \\
  $B_{sJ}(6114)$      &$6114\pm3\pm5$                       &$66\pm18\pm21$        & $0(?^?)$\cite{LHCb:2020pet}     \\
    $B_{sJ}(6109)$      &$6108.8\pm1.1\pm0.7$                 &$22\pm5\pm4$          & $0(?^?)$\cite{LHCb:2020pet}     \\
  $B_{sJ}(6158)$      &$6158\pm4\pm5$                       &$72\pm18\pm25$        & $0(?^?)$\cite{LHCb:2020pet}     \\
  \hline\hline

\end{tabular}}
\end{center}
\end{table}

In this paper, we use the screened nonrelativistic quark model and the $^3P_0$ model to study the spectrum and the strong decay properties of the $B_s$ mesons, and also to explore the possible assignments of the two resonances recently observed by the LHCb Collaboration.

This article is organized as follows. In Sec.~\ref{sec:model}, we give a brief introduction about the screened nonrelativestic quark model and the $^3P_0$ model. In Sec.~\ref{sec:result}, the numerical results and the discussions are presented. Finally, the summary is given in Sec.~\ref{sec:summary}.

\section{Theoretical models}
\label{sec:model}

\subsection{Screened nonrelativistic quark model}
The nonrelativistic quark model mainly includes the confinement term , the spin-dependent term, and the one-loop correction for the spin-dependent terms~~\cite{Gupta:1981pd,Pantaleone:1985uf,Lakhina:2006fy}, and the Hamiltonian for a $q\bar{q}$ meson system is defined as~\citep{Li:2010vx,Lu:2016bbk},
    \begin{equation}
    \label{ha}
   \mathcal{H} = \mathcal{H}_0+\mathcal{H}_{sd}+C_{q\bar{q}}, 
    \end{equation}
where $\mathcal{H}_0$ is the zeroth-order Hamiltonian, $\mathcal{H}_{sd}$ is the spin-dependent Hamiltonian, and $C_{q\bar{q}}$ is a constant, which will be fixed to experimental data.
The $\mathcal{H}_{0}$ can be compressed as,    \begin{eqnarray}
    \mathcal{H}_0 &=& \frac{\boldsymbol{p}^2}{M_r}-\frac{4}{3}\frac{{\alpha}_s}{r}+br+\frac{32{\alpha}_s{\sigma}^3 e^{-{\sigma}^2r^2}}{9\sqrt{\pi}m_qm_{\bar{q}}} {\boldsymbol{S}}_{q} \cdot {\boldsymbol{S}}_{\bar{q}}, \label{eq:H0}
     \end{eqnarray}
where the confinement interaction includes the standard Coulomb potential $-4\alpha_s/3r$ and the linear scalar potential $br$. The last term is the hyperfine interaction that could be treated nonperturbatively.
 $\boldsymbol{p}$ is quark momentum in the system of $q\bar{q}$ meson, $r=|\vec{r}\,|$ is the $q\bar{q}$ separation, $M_r=2m_qm_{\bar{q}}/(m_q+m_{\bar{q}})$, $m_q$ ($m_{\bar{q}}$) and $\boldsymbol{S}_{q}$ (${\boldsymbol{S}}_{\bar{q}}$) are the reduced mass of the $q\bar{q}$ system, the mass and spin of the constituent quark $q$ (antiquark $\bar{q}$), respectively.

The spin-dependent term $\mathcal{H}_{sd}$ is,
    \begin{eqnarray}
      \mathcal{H}_{sd} &=& \left(\frac{\boldsymbol{S}_{q}}{2m_q^2}+\frac{{\boldsymbol{S}}_{\bar{q}}}{2m_{\bar{q}}^2}\right) \cdot \boldsymbol{L}\,\left(\frac{1}{r}\frac{dV_c}{dr}+\frac{2}{r}\frac{dV_1}{dr}\right)\nonumber\\
      &&+\frac{{\boldsymbol{S}}_+ \cdot \boldsymbol{L}}{m_qm_{\bar{q}}}\left(\frac{1}{r} \frac{dV_2}{r}\right) \nonumber\\
      && +\frac{3{\boldsymbol{S}}_{q} \cdot \hat{\boldsymbol{r}}\,{\boldsymbol{S}}_{\bar{q}} \cdot \hat{\boldsymbol{r}}-{\boldsymbol{S}}_{q} \cdot {\boldsymbol{S}}_{\bar{q}}}{3m_qm_{\bar{q}}}V_3\nonumber\\
      && +\left[\left(\frac{{\boldsymbol{S}}_{q}}{m_q^2}-\frac{{\boldsymbol{S}}_{\bar{q}}}{m_{\bar{q}}^2}\right)+\frac{{\boldsymbol{S}}_-}{m_qm_{\bar{q}}}\right] \cdot \boldsymbol{L} V_4,
\end{eqnarray}
with
\begin{eqnarray}
  V_c &=& -\frac{4}{3}\frac{{\alpha}_s}{r}+br,\nonumber \\
  V_1 &=& -br-\frac{2}{9\pi}\frac{{\alpha}_s^2}{r}\left[9\,{\rm ln}(\sqrt{m_qm_{\bar{q}}}r)+9{\gamma}_E-4\right],\nonumber\\
  V_2 &=& -\frac{4}{3}\frac{{\alpha}_s}{r}-\frac{1}{9\pi}\frac{{\alpha}_s^2}{r}\left[-18\,{\rm ln}(\sqrt{m_qm_{\bar{q}}}r)+54\,{\rm ln}(\mu r) \right.\nonumber\\
  &&\left. +36{\gamma}_E+29\right],\nonumber\\
  V_3 &=& -\frac{4{\alpha}_s}{r^3}-\frac{1}{3\pi}\frac{{\alpha}_s^2}{r^3}\left[-36\,{\rm ln}(\sqrt{m_qm_{\bar{q}}}r)+54\,{\rm ln}(\mu r) \right.\nonumber\\
  &&\left. +18{\gamma}_E+31\right],\nonumber\\
  V_4 &=& \frac{1}{\pi}\frac{{\alpha}_s^2}{r^3}{\rm ln}\left(\frac{m_{\bar{q}}}{m_q}\right),
\end{eqnarray}
where $\boldsymbol{S}_{\pm}={\boldsymbol{S}}_q\pm{\boldsymbol{S}}_{\bar{q}}$, $\boldsymbol{L}$ is the relative orbital angular momentum of the $q\bar{q}$ system. We take Euler constant $\gamma_E=0.5772$, the scalar $\mu=1$~GeV, ${\alpha}_s=0.53$, $b=0.135$~GeV$^2$, $\sigma=1.13$~GeV, $m_u=m_d=0.45$~GeV, $m_s=0.55$~GeV and $m_b=4.5$~GeV \citep{Lakhina:2006fy}.

The screening effects are introduced by the following  replacement,
\begin{eqnarray}
br\to V^{\text{scr}}(r)=\frac{b(1-e^{-\beta r})}{\beta}, \label{eq:screened}
\end{eqnarray}
where $V^{\text{scr}}(r)$ behaves like $br$ at short distances and constant $b/\beta$ at large distance~\cite{Song:2015nia, Song:2015fha}, $\beta$ is the parameter which  controls the power of the screening effects. 
One can find that the screened potential approximates the liner potential for a small distance $r$, and will be softened for large distance $r$.
Since the distance between the quarks in the excited bottom-strange mesons is larger than the one of the ground bottom-strange meson, it is expected that the spectrum of the screened potential model is more sensitive  for the excited bottom-strange mesons.

The spin-orbit term $\mathcal{H}_{sd}$ can be decomposed into
symmetric part $\mathcal{H}_{sym}$ and antisymmetric part $\mathcal{H}_{anti}$, which  can be expressed as \citep{Lu:2016bbk}
\begin{eqnarray}
\mathcal{H}_{sym} &=& \frac{{\boldsymbol{S}}_+ \cdot {\boldsymbol{L}}}{2}\left[\left(\frac{1}{2m_q^2}+\frac{1}{2m_{\bar{q}}^2}\right) \left(\frac{1}{r}\frac{dV_c}{dr}+\frac{2}{r}\frac{dV_1}{dr}\right)\right. \nonumber \\
&& \left.+\frac{2}{m_qm_{\bar{q}}}\left(\frac{1}{r} \frac{dV_2}{r}\right)+\left(\frac{1}{m_q^2}-\frac{1}{m_{\bar{q}}^2}\right)V_4\right],
\end{eqnarray}
\begin{eqnarray}
\mathcal{H}_{anti} &=& \frac{{\boldsymbol{S}}_- \cdot {\boldsymbol{L}}}{2}\left[\left(\frac{1}{2m_q^2}-\frac{1}{2m_{\bar{q}}^2}\right) \left(\frac{1}{r}\frac{dV_c}{dr}+\frac{2}{r}\frac{dV_1}{dr}\right)\right. \nonumber \\
&& \left.+\left(\frac{1}{m_q^2}+\frac{1}{m_{\bar{q}}^2}+\frac{2}{m_qm_{\bar{q}}}\right)V_4\right].
\end{eqnarray}
The antisymmetric part $\mathcal{H}_{anti}$ gives rise to the
the spin-orbit mixing of the heavy-light mesons with different total spins but with the same total angular momentum, such as
$B_s(n{}^3L_L)$ and $B_s(n{}^1L_L)$.  Hence, the mixing of the two physical states $B_{sL}(nL)$ and $B_{sL}^\prime(nL)$ can be
expressed as,
\begin{equation}
\left(
\begin{array}{cr}
B_{sL}(nL)\\
B^\prime_{sL}(nL)
\end{array}
\right)
 =\left(
 \begin{array}{cr}
\cos \theta_{nL} & \sin \theta_{nL} \\
-\sin \theta_{nL} & \cos \theta_{nL}
\end{array}
\right)
\left(\begin{array}{cr}
B_s(n^1L_L)\\
B_s(n^3L_L)
\end{array}
\right),
\label{Bmixing1}
\end{equation}
where the $\theta_{nL}$ is the mixing angles.

With above formalism, one can solve the Schr\"{o}dinger equation with Hamiltonian $\mathcal{H}$ of Eq.~(\ref{ha}) to obtain the mass spectrum and the  meson wave functions, where the wave functions will be used to calculate the strong decays of excited bottom-strange mesons in the $^3P_0$ model.

\subsection{The $^3P_0$ model}
The $^3P_0$ model was proposed by Micu \cite{Micu:1968mk} and further developed by Le Yaouanc
\cite{LeYaouanc:1972vsx, LeYaouanc:1974cvx, LeYaouanc:1977fsz,LeYaouanc:1977gm},  and it has been widely used to calculate the OZI allowed decay processes \cite{Roberts:1992esl,Barnes:1996ff,Barnes:2002mu,
Close:2005se,Barnes:2005pb,
Li:2009rka,Li:2009qu,Li:2010vx,Lu:2014zua,Pan:2016bac,Lu:2016bbk,Hao:2019fjg,Feng:2022esz,Wang:2018rjg,Wang:2019mhs,Song:2015nia,Song:2015fha,Xue:2018jvi,Wang:2017pxm}.
In this model, the meson decay occurs through the regroupment between the $q\bar{q}$ of the initial meson and the another $q\bar{q}$ pair created from vacuum with the quantum numbers $J^{PC}=0^{++}$. The transition operator $\mathcal{T}$ of the decay  $A\rightarrow BC$ in the $^3P_0$ model is given by
\begin{eqnarray}
&\mathcal{T}=-3\gamma \sum \limits_{m} \langle1m;1-m|00\rangle \int d^3\boldsymbol{p}_3 d^3\boldsymbol{p}_4 \delta^3(\boldsymbol{p}_3+\boldsymbol{p}_4)\nonumber\\
&\mathcal{Y}_{1m}\left(\frac{\boldsymbol{p}_3-\boldsymbol{p}_4}{2}\right) \chi^{34}_{1,-m} \phi^{34}_{0} \left(\omega^{34}_{0}\right)b^{\dag}_{3}(\boldsymbol{p}_3) d^{\dag}_{4}(\boldsymbol{p}_4),
\end{eqnarray}
where ${\cal{Y}}^m_1(\boldsymbol{p})\equiv|\boldsymbol{p}|^1Y^m_1(\theta_p,\phi_p)$ is solid harmonic polynomial in the momentum space of the created quark-antiquark pair.
$\chi^{34}_{1, -m}$, $\phi^{34}_0$, and $\omega^{34}_0$
 are the spin, flavor, and color wave functions, respectively.
The paramtere $\gamma$ is the quark pair creation strength parameter  for $u \bar{u}$ and $d \bar{d}$ pairs, and for $s\bar{s}$ we take $\gamma_{s\bar{s}}=\gamma\frac{m_u}{m_s}$ \cite{LeYaouanc:1977gm}. The parameter $\gamma$ can be determined by fitting to the experimental data. 
The partial wave amplitude ${\cal{M}}^{LS}(\boldsymbol{P})$ of the decay  $A\rightarrow BC$ is be given by,
\begin{eqnarray}
    {\cal{M}}^{LS}(\boldsymbol{P})&=&
    \sum_{ M_{J_B},M_{J_C}, M_S,M_L}
    \langle LM_LSM_S|J_AM_{J_A}\rangle \nonumber\\
    &&\langle
    J_BM_{J_B}J_CM_{J_C}|SM_S\rangle\nonumber\\
    &&\int
    d\Omega\,\mbox{}Y^\ast_{LM_L} 
  {\cal{M}}^{M_{J_A}M_{J_B}M_{J_C}}
    (\boldsymbol{P}),
\end{eqnarray}
where ${\cal{M}}^{M_{J_A}M_{J_B}M_{J_C}}
(\boldsymbol{P})$ is the helicity amplitude,
\begin{eqnarray}
\langle
BC|T|A\rangle =\delta^3(\boldsymbol{P}_A-\boldsymbol{P}_B-\boldsymbol{P}_C)\nonumber\\
               {\cal{M}}^{M_{J_A}M_{J_B}M_{J_C}}(\boldsymbol{P}).
\end{eqnarray}
Here, $|A\rangle$, $|B\rangle$, and $|C\rangle$ denote the mock meson states.
Then, the decay width
$\Gamma(A\rightarrow BC)$ can be expressed as

\begin{eqnarray}
\Gamma(A\rightarrow BC)= \frac{\pi
P}{4M^2_A}\sum_{LS}|{\cal{M}}^{LS}(\boldsymbol{P})|^2,
\end{eqnarray}
where $P=|\boldsymbol{P}|=\frac{\sqrt{[M^2_A-(M_B+M_C)^2][M^2_A-(M_B-M_C)^2]}}{2M_A}$,
 $M_A$, $M_B$, and $M_C$ are the masses of the mesons $A$, $B$,
and $C$, respectively. The spatial wave functions of the mesons in the $^3P_0$ model are obtained by solving the Schr$\ddot{o}$dinger equation in Eq. (\ref{ha}).

\section{Results and discussions}
\label{sec:result}

In the calculation, the screened parameter $\beta=0.025$~GeV and the constant $C_{q\bar{q}}=0.1035$~GeV were obtained by fitting the well-known states $B_s(1^1S_0)$, $B_s^*(1^3S_1)$, and $B_{s2}^*(5840)(1^3P_2)$. The other parameters are taken from Ref.~\cite{Lakhina:2006fy}. The $^3P_0$ model parameter $\gamma=0.354$ is obtained by fitting the total decay width of the $B_{s2}^*(5840)$, which is regarded as the $B_s(1^3P_2)$. With these parameters, the predicted ratio of the $B_{s2}^*(5840)$ decay modes,
\begin{equation}
\frac{\Gamma(B_{s2}^*(5840)\to B^{*+}K^-)}{\Gamma(B_{s2}^*(5840)\to B^+K^-)} = 0.095,
\end{equation} which is consistent with LHCb experimental data of $0.093\pm0.013\pm0.012$~\cite{LHCb:2012iuq}.

The predicted masses of the $B_s$ mesons are listed in Table~\ref{mass}, where we also show the predictions of other theoretical works for comparison. The $B_s$[$B_s(1^1S_0)$], $B_s^*$[$B_s(1^3S_1)$], and $B_{s2}^*(5840)$[$B_s(1^3P_2)$] can be well described in the spectrum. The masses of the $B_s(1^3D_1)$ and $B_s(1^3D_3)$  are predicted to be 6061~MeV and 6117~MeV, in good agreement with  the experimental results of the $B_{sJ}(6064)$ ($6063.5\pm1.2\pm0.8$~MeV) and $B_{sJ}(6114)$  ($6114\pm3\pm5$~MeV), respectively, which indicates that the two states could be the possible candidates of the $B_s(1^3D_1)$ and $B_s(1^3D_3)$ .

Of course, only the mass information is not enough to establish these assignments. We also calculate the strong decay widths of the $B_s$ mesons, as shown in Table~\ref{decay}.
One can find that the predicted width of 23~MeV for $B_s(1^3D_3)$ is in good agreement with the measured width  $26\pm4\pm4$~MeV  of the $B_{sJ}(6064)$, which supports the $B_s(1^3D_3)$ assignment of  the $B_{sJ}(6064)$. In addition, the predicted width for $B_s(1^3D_1)$ is 127~MeV, reasonably consistent with the one of $B_{sJ}(6114)$ if taking into account the large experimental uncertainties. Thus, the $B_{sJ}(6114)$ could be explained as the $B_s(1^3D_1)$ state, and more precise measurements will be helpful to pin down this assignment. 


As we discussed in the introduction, if a decay through $B^{*\pm}K^\mp$ with a missing photon from the $B^{*\pm} \to B^\pm\gamma$ decay is assumed, the masses and widths of the two states observed by LHCb will shift, and the two states are named as $ B_{sJ}(6109)$ and $B_{sJ}(6158)$~\cite{LHCb:2020pet}.
In this case, the mass and width of the $B_{sJ}(6109)$ are close to the predicted mass (6132~MeV) and width (43~MeV) of $B_{s2}^\prime(1D)$, which implies that $B_{sJ}(6109)$ could be regarded as the $B_{s2}^\prime(1D)$ state. 
On the other hand, the mass and width of the $B_{sJ}(6158)$ are close to the predicted mass (6194~MeV) and width (75~MeV) of $B_{s1}(2P)$, respectively, which supports the assignment of the  $B_{sJ}(6158)$ as the $B_{s1}(2P)$ state. It should be stressed that the two solutions could be not distinguished according to the present LHCb measurements. Thus, the future precise measurements of their masses, widths, and the quantum numbers of the spin-parity would be helpful to shed light on this problem, and deepen our understanding the spectra of the bottom-strange mesons.

\begin{table*}[!htbp]
\begin{center}
\caption{ \label{mass} The mass spectrum of the $B_s$ mesons predicted by the screened quark model in units of MeV. The mixing angles of $B_{sL}-B^\prime_{sL}$ calculated in
this work are $\theta_{1P} =-55.8^\circ$, $\theta_{2P} =-54.3^\circ$ and $\theta_{1D}=-50.3^\circ$.}
\footnotesize
\begin{tabular}{lccccccccccc}
\hline\hline
  State                 &PDG                     & Ours     &NR\cite{Lu:2016bbk}    &GI\cite{Godfrey:2016nwn}     &DRV\cite{Ebert:2009ua}    &KA\cite{Gandhi:2022nnk} &VRA\cite{Patel:2022hhl}  \\\hline
  $B_s(1^1S_0)$         &$5366.92\pm0.10$        &5367      &5362   &5394   &5372   &       &5359      \\
  $B_s(1^3S_1)$         &$5415.4^{+1.8}_{-1.5}$  &5419      &5413   &5450   &5414   &       &5415      \\
  $B_s(2^1S_0)$         &                        &5947      &5977   &5984   &5976   &6025   &5980          \\
  $B_s(2^3S_1)$         &                        &5972      &6003   &6012   &5992   &6033   &5993          \\
  $B_s(1^3P_0)$         &                        &5753      &5756   &5831   &5833   &5709   &5798          \\
  $B_{s1}(1P)$          &                        &5797      &5801   &5857   &5831   &5768   &5818          \\
  $B^\prime_{s1}(1P)$   & $5828.70\pm 0.20$      &5825      &5836   &5861   &5865   &5875   &5846          \\
  $B_s(1^3P_2)$         &$5839.86\pm0.12$        &5840      &5851   &5876   &5842   &5890   &5838          \\
  $B_s(2^3P_0)$         &                        &6158      &6203   &6279   &6318   &6387   &6292          \\
  $B_{s1}(2P)$          &$6158\pm4\pm5$          &6194      &6241   &6279   &6321   &6393   &6304          \\
  $B^\prime_{s1}(2P)$   &                        &6236      &6297   &6296   &6345   &6470   &6320          \\
  $B_s(2^3P_2)$         &                        &6249      &6309   &6295   &6359   &6476   &6316          \\
  $B_s(1^3D_1)$         &$6114\pm3\pm5$          &6117      &6142   &6182   &6209   &6247   &6144          \\
  $B_{s2}(1D)$          &                        &6053      &6087   &6169   &6189   &6256   &6139          \\
  $B^\prime_{s2}(1D)$   &$6108.8\pm1.1\pm0.7$    &6132      &6159   &6196   &6218   &6292   &6135          \\
  $B_s(1^3D_3)$         &$6063.5\pm1.2\pm0.8$    &6061      &6096   &6179   &6191   &6297   &6139          \\
  \hline\hline

\end{tabular}
\end{center}
\end{table*}

\begin{table*}[!htbp]
\begin{center}
\caption{ \label{decay} The decay widths of the $B_s$ mesons in units of MeV. The label `$-$' means that the channel is forbidden or there is no  experimental information.}
\footnotesize
\begin{tabular}{lccccccccccc}
\hline\hline
   State               &  &$BK$ &$B^*K$ &$BK^*$ &$B^*K^*$ &$B_s\eta$  &$B_s^*\eta$  &Total &Exp.\\\hline
  $B_{s1}(1P)$         &$-$                    &$-$    &$-$     &$-$      &$-$      &$-$   &$-$  &$-$    &$-$    \\
  $B^\prime_{s1}(1P)$  &$B_{s1}(5830)$       &$-$    &0.04    &$-$      &$-$      &$-$   &$-$  &0.03   &$0.5\pm0.3\pm 0.3$      \\ 
  $B_s(1^3P_2)$        &$B_{s2}^*(5840)$     &1.4    &0.1     &$-$      &$-$      &$-$   &$-$  &1.5    &$1.49\pm0.27$      \\   
  $B_s(2^3P_0)$        &$-$                    &61.5   &$-$     &$-$      &$-$      &5.3   &$-$  &66.8   &$-$   \\
  $B_{s1}(2P)$         &$B_{sJ}(6158)$          &$-$    &58.2    &13.2     &$-$      &$-$   &3.9 &75.3   &$72\pm18\pm25$    \\
  $B^\prime_{s1}(2P)$  &$-$                     &$-$    &15.0    &41.8     &52.4     &$-$   &4.8  &113.9   &$-$    \\
  $B_s(2^3P_2)$        &$-$                    &1.7    &7.3     &18.7     &139.7    &1.3   &2.8  &171.6  &$-$ \\
  $B_s(1^3D_1)$        &$B_{sJ}(6114)$       &67.5   &37.4    &$-$      &$-$      &15.3  &6.8  &127.0  &$66\pm18\pm21$       \\
  $B_{s2}(1D)$         &$-$                    &$-$    &115.5    &$-$      &$-$      &$-$   &11.7 &127.3   &$-$    \\
  $B^\prime_{s2}(1D)$  &$B_{sJ}(6109)$                    &$-$    &41.8    &$-$      &$-$      &$-$   &1.2  &43.0   &$22\pm5\pm4$    \\
  $B_s(1^3D_3)$        &$B_{sJ}(6064)$       &12.0   &10.6    &$-$      &$-$      &0.3   &0.1  &23.0   &$26\pm4\pm4$      \\
\hline\hline
\end{tabular}
\end{center}
\end{table*}

\begin{figure}[!htpb]
  \centering
   \begin{tabular}{c}
  \includegraphics[scale=0.7]{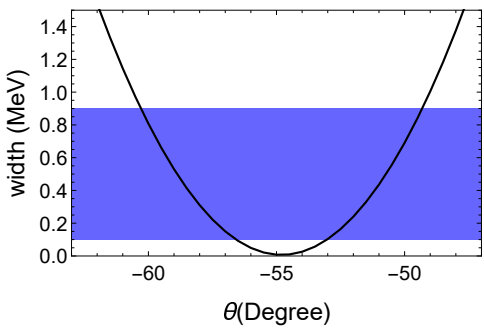}
   \end{tabular}
  \caption{Total decay widths of the $B_{s1}(5830)$ as the $1P^\prime$ versus the mixing angle. The blue band denotes the uncertainties of the experimental data. }
  \label{fig3}
\end{figure}

According to the predicted mass spectrum of Table~\ref{mass}, the predicted mass 5840~MeV of the $B_{s1}^\prime(1P)$ is very close to the one of the $B_{s1}(5830)$. With the mixing angle of $-55.8^\circ$, the decay widths of the $B_{s1}^\prime(1P)$ are calculated, as shown in Table~\ref{decay}, and the predicted total decay width is very small, in good agreement with the experimental value $0.5\pm 0.3\pm 0.3$~MeV of the $B_{s1}(5830)$. We show the total decay widths of the $B_{s1}^\prime(1P)$ versus the mixing angle in Fig.~\ref{fig3}, where one can find the total decay width is still consistent with the experimental data with the mixing angle in the range of $-60^\circ \sim -50^\circ$.



In addition, we also predict the decay widths of the other excited $B_s$ mesons with the predicted masses of Table~\ref{mass}. For the $B_s(2^3P_0)$, its total decal width is predicted to be 66.8~MeV, and the dominant decay mode is $BK$ with the branching fraction 92\%.
The total decay width of   $B'_{s1}(2P)$ is  113.9~MeV, while the dominant decay modes are $BK^*/B^*K^*$. The predicted width and the dominant decay mode of the $B_{s}(2^3P_2)$ are 171.6~MeV and $B^*K^*$, respectively. In addition, the decay widths of the $B_{s2}(1D)$ are 127.3~MeV, and the dominant decay mode is $B^*K$.
Our results should be helpful to search for them in experiments, such as LHC.

\section{Summary}
\label{sec:summary}
Recently, the LHCb Collaboration has observed two resonances $B_{sJ}(6064)$ and $B_{sJ}(6114)$ assuming they decay directly to the $BK$ final states. However, their masses and widths will shift if a decay through $B^*K$ with a missing photon from the $B^*$ decay, and those two resonances are named as $B_{sJ}(6109)$ and $B_{sJ}(6158)$.

Motivated by the recently LHCb measurements, in this paper we calculate the spectrum of the $B_s$ mesons within the screened nonrelativistic quark model, and  also investigate the strong decay properties of these mesons with  the $^3P_0$ model.

By comparing with the experimental data, it is found that the $B_{sJ}(6064)$ and $B_{sJ}(6114)$ states, as the first solution of the recently LHCb measurements, could be explained as the $B_s(1^3D_3)$ and $B_s(1^3D_1)$, respectively. In addition, the $B_{sJ}(6109)$ and $B_{sJ}(6158)$ states, as the second solution of the LHCb measurements, could be explained as the $B'_{s2}(1D)$ 
and $B_{s1}(2P)$, respectively. Since those two solutions could be not distinguished according to the present LHCb measurements, thus the future precise measurements of their masses, widths, and the quantum numbers of the spin-parity would be helpful to shed light on this problem, and deepen our understanding the spectra of the bottom-strange mesons.

In addition, we suggest that the state $B_{s1}(5830)$ could be explained as the candidate of the $B_{s1}^\prime(1P)$ state.  
With the predicted masses of the other excited $B_s$ states, we also predict their decay widths as well as the dominant decay modes,  which should be helpful for experiments to search for them.

It should be stressed that there are already many theoretical studies about the family of the bottom-strange mesons. Comparing with those works, we have adopted the nonrelativistic quark model by taking into account the screening effects, which play an important role for the higher radial and orbital excited mesons, and our results could give better descriptions for all the existed bottom-strange mesons.

\begin{acknowledgements}
This work is partly supported by the Natural Science Foundation of Henan under Grand Nos. 222300420554 and 232300421140, the Project of Youth Backbone Teachers of Colleges and Universities of Henan Province (2020GGJS017), the Youth Talent Support Project of Henan (2021HYTP002), and the Open Project of Guangxi Key Laboratory of Nuclear Physics and Nuclear Technology, No.NLK2021-08.
\end{acknowledgements}

\bibliographystyle{unsrt}
\bibliography{cite}  

\end{document}